\documentclass{elsart}
\usepackage{graphics,natbib,graphicx,psfig,amssymb,ccaption}
\journal{Journal of Scientometric Research}
\input psfig.sty
\begin{document}
\begin{frontmatter}
\title{A New Ranking Scheme for the Institutional Scientific Performance}
\author[istanbul1]{S. Bilir\corauthref{cor}},
\corauth[cor]{Corresponding author. Tel.: +90-212-440 00 00-10534}
\ead{sbilir@istanbul.edu.tr}
\author[sabanci]{ E. G{\"o}{\u g}{\"u}{\c s}},
\author[istanbul2]{{\"O}. {\"O}nal Ta{\c s}}, and 
\author[istanbul2]{T. Yontan}
\address[istanbul1]{Istanbul University, Faculty of Science, Department 
of Astronomy and Space Sciences, 34119, Istanbul, Turkey}
\address[sabanci]{Sabanc\i~University, Faculty of Engineering and Natural
  Sciences, 34956, Orhanl\i-Tuzla, Istanbul, Turkey}
\address[istanbul2]{Istanbul University, Graduate School of Science and 
  Engineering, Department of Astronomy and Space Sciences, 34116, Istanbul, 
  Turkey}

\begin{abstract}
We propose a new performance indicator to evaluate the productivity of research
institutions by their disseminated scientific papers. The new quality measure
includes two principle components: the normalized impact factor of the journal
in which paper was published, and the number of citations received per year
since it was published. In both components, the scientific impacts are weighted
by the contribution of authors from the evaluated institution. As a whole, our
new metric, namely, the institutional performance score takes into account both
journal based impact and articles specific impacts. We apply this new scheme to 
evaluate research output performance of Turkish institutions specialized in 
astronomy and astrophysics in the period of 1998-2012. We discuss the implications 
of the new metric, and emphasize the benefits of it along with comparison to other
proposed institutional performance indicators.
\end{abstract}

\end{frontmatter}

\section{Introduction}
A natural extension of evaluating the research performance of individual 
scientists is to evaluate the research output productivity of research 
institutes. This is, however, a more challenging task than assessing the 
output records of an individual scientist for various reasons. First of 
all, the number of scientists affiliated varies remarkably between 
institutes. This is easily handled in the evaluation of the research 
performance of an institute by normalizing the research outputs with the 
number of affiliated researchers. Another important factor is the impact 
of the research output. At this front, the {\em h}-index \citep{Hirsch05} 
and some of its variants \citep{Braun06, Egghe06, VanRaan06, Jin07, Guan08, 
Vanclay08, Schreiber11} are usually employed.

Along with the wider use of advanced technology and methodologies in 
scientific research, the nature of research teams is also evolving. 
Unlike a few decades ago, scientific investigations performed by teams of 
about 10 scientists or more are not uncommon. The size of research teams 
in some cases can be as large as hundreds, such as, the Large Hadron 
Collider collaboration at CERN\footnote{http://lhcb-public.web.cern.ch/lhcb-public/}, 
which includes scientists affiliated with many different institutions. In the 
dissemination of these scientific efforts (most commonly in the form of 
research articles), the contribution of each team member (that is, co-author) 
is not usually reported explicitly. Therefore, it would not be a fair evaluation 
of the respective institutions when these large collaboration articles are 
assessed without author contributions are taken into considerations.
To account for authorship credit, various ways were proposed, such as, 
the harmonic author credit \citep{Hagen08} and the i$^{th}$ author credit 
\citep{Liu12}, both of which credits the author based on the rank in 
the author list, or the fractional author credit \citep{Liu12} which 
credits all authors equally.

There have been numerous extensive studies for the scientific productivity 
evaluations of research institutions. {\citet{Vieira10} investigated 
research impact for scientific institutions using an indicator that includes 
the paper productivity as well as their citation performance. \citet{Batista06}
proposed a measure that is interrelated to the {\em h}-index: They introduced $h_I$ 
which is the ratio of the square of {\em h}-index of the institutional papers to the 
number of authors of these articles. \citet{Abramo13} derived an indicator which is 
obtained by normalizing the institutional {\em h}-index with the number of full time 
research personnel of the institute. \citet{Franceschini10} suggested a structured 
technique to evaluate scientific output of research groups, in which they employ 
{\em h}-index as the key ingredient. Recently, \citet{Franceschini13} proposed the 
success index for evaluating research institutions which primarily takes into account 
institutional papers with greater citation records. \citet{Boell10} proposed a ranking 
scheme based on the square of the journal impact factors. Note the fact that these 
performance indicators do not completely involve the effects of all the 
above-mentioned factors, in particular the author contribution to the impact of
scientific output.

Here, we propose a new evaluation scheme for the institutional research 
productivity, that takes into account the scientific impact of the output as 
well as the extend of the scientific contribution of its researchers. We 
introduce this new institutional performance score scheme in the next section. 
Then, we apply our proposed performance indicator method to the research outputs 
of institutions performing research in astronomy and astrophysics in Turkey, 
based on their article productivity between 1998 and 2012. Finally, we discuss 
the implications of our results, and compare the results obtained with this new 
scheme and through other techniques in Section 4. 

\section{Material and methods}

In order to obtain the complete dataset for astronomy and astrophysics research papers, 
we used Thomson Reuters Web of Knowledge\footnote{http://apps.webofknowledge.com}, which 
includes 12 different databases of single and interdisciplinary citation indices. This 
database contains the list of all journals covered in Science Citation Index (SCI) and 
provides the citation counts without self citations for individual papers since 1980 to 
present day. In the database, we have identified 1702 publications in ``Astronomy and 
Astrophysics'' whose authors or co-authors were based in Turkey and published in 56 SCI 
journals in the period from 1980 to 2012. After excluding papers with overlapping fields, 
such as physics particles fields, geosciences multidisciplinary, meteorology atmospheric 
sciences, engineering aerospace, geochemistry geophysics, mathematics interdisciplinary 
applications and remote sensing, the total number of publications was reduced to 1062. 
According to document types, these 1062 papers were divided into seven 
groups: articles (976), proceedings (37), letters (16), reviews (15), errata (10), 
research notes (7) and editorial notes (1). It was also found that 37 of these studies 
were presented at meetings before they were published, 10 of them were corrected then 
re-published. Due to these reasons, we only considered the articles, letters, reviews, 
and editorial notes, which resulted in sample size of 1015 publications. In Fig. 1, we 
present the distribution of these 1015 publications over time. Note that 782 of these 
papers had the leading author from Turkish institutions, while in 233 papers; the 
leading authors were from international institutions. 

\begin{figure}
\centering
   \includegraphics[scale=0.55]{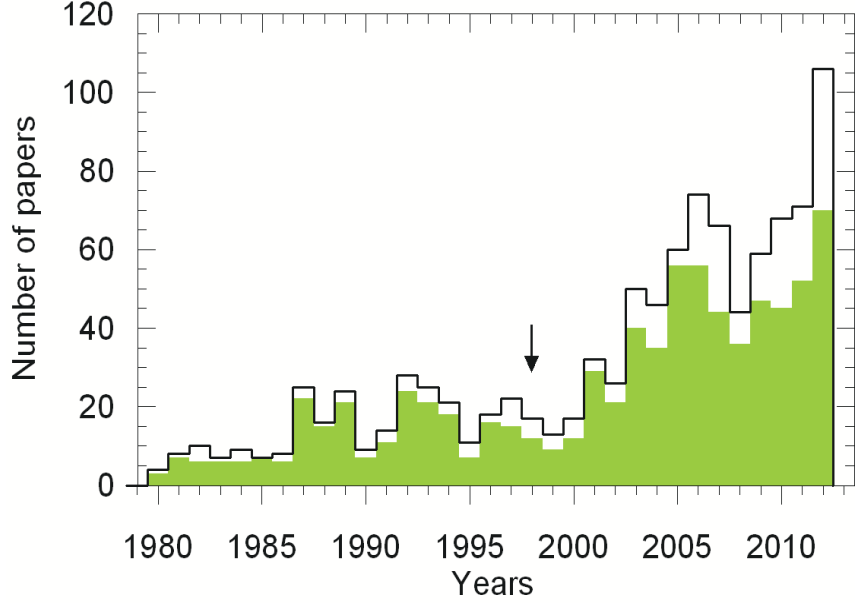}
   \caption{Distribution of research papers in astronomy and astrophysics which were 
published in SCI journals from 1980 to 2012. Green histograms represent the same 
distribution for the papers with the leading authors from institutions in Turkey.}
\end{figure}

It can be seen from Fig. 1 that there is a prominent increment in number of 
publications that were published in SCI journals starting from about 2000. 
\citet{Bilir13} suggested that this increase in scientific productivity was 
motivated by the application of the improved academic assignment criteria in 
1998, common use of Internet, and larger scale research opportunities provided 
by scientific research units of the universities. Therefore, in our study here 
we analyzed research papers that published between the years of 1998 and the 
end of 2012. We should note the important fact that the number of citations 
for each publication was determined as of 31 August 2013. This way, even the 
latest publications had about one year of visibility, since papers are 
usually published in SCI journals can take a couple of months before they are 
listed on Web of Knowledge.

Astronomy and astrophysics studies in Turkey are currently conducted in three 
main departments such as Astronomy and Space Sciences in Ankara, Ege, Erciyes 
and \.Istanbul universities; Astronomy and Space Technologies in Akdeniz, 
{\c C}anakkale Onsekiz Mart (COMU) universities and Physics in 
Bo{\u g}azi{\c c}i (BOUN), {\c C}ukurova, \.Istanbul K\"ult\"ur, Middle East 
Technical (METU) and Sabanc\i~universities.

Note that astrophysical research has also been conducted at various sub 
divisions of the Turkish Scientific and Technological Research Council 
(T{\"U}B{\.I}TAK)\footnote{www.tubitak.gov.tr}. We have identified 749 papers 
that researchers based in Turkey have been leading authors or co-authors and 
have been published in SCI journals between 1998 and 2012. Researchers from 48 
institutions contributed to these 749 publications. When we consider the papers 
with leading authors based in Turkey, the number of papers published within 
the same time frame reduces to 564, which were contributed by researchers from 
37 institutions inside Turkey. 

\section{Theory/calculation}
As we have outlined in the first section, there is currently no quality 
indicator to rank scientific productivity of research institutions that takes 
into account scientific impact and author contribution. We introduce below our 
article productivity based new ranking scheme, which we call the institutional 
performance score ({\em IPS}), which consists of two additive terms: ({\em i}) impact 
factor ($IF$) of journal for the year that the article has been published 
multiplied with the contribution of each co-author ($AC$) to the institutional 
article, and ({\em ii}) the ratio of the number of citations received ($n_{citations}$) 
to the number of years passed since the paper has been published ($n_{years}$),
and also multiplied by $AC$.

\begin{equation}
IPS=\frac{1}{N}\sum_{i=1}^{N} \left ((IF)_{i} +\frac{n_{citations, i}}{n_{years, i}} \right) \times (AC)_{i},
\end{equation}
where $N$ is the total number of institutional articles published. In this scheme, 
the $AC$ parameter is simply the ratio of the number of co-authors from a particular 
institute to the total number of co-authors. For example, if an article is published 
by five researchers; three of them are from Institute A and two of them from 
Institute B, then the author contribution of this paper to Institute A is 3/5 and 
that to Institute B is 2/5. The latter term represents the scientific impact of 
an article, which diminishes over time if it is not cited at a steady pace. Effectively,
this indicator combines author contribution added impact gained by the journal in which
a particular article was published, and by the article itself. 

\section{Results}

We apply our proposed institutional performance indicator to the Turkish institutions 
performing research in astronomy and astrophysics, and disseminate their outputs in 
the form of scientific articles. Note that the impact factor of the journal in the year 
that a paper is published is one of the essential inputs for our new performance indicator 
definition. For this purpose, the impact factors for the nine mostly preferred  
SCI journals between 1998 and 2012 were compiled and presented in Table 1. In the bottom 
row of Table 1, we provide the 15-year averages of annual impact factors for each of these 
nine journals. 

\begin{table}
\setlength{\tabcolsep}{5pt}
\renewcommand{\arraystretch}{1}
\caption{ Impact factors of SCI journals in astronomy and astrophysics between 1998 and 2012.}
\begin{tabular}{cccccccccc}
\hline
Years& MNRAS &  A\&A &        ApJ &       NewA &      IJMPD &         AN &     Ap\&SS &       PASA &         AJ \\
\hline
1998 & 3.960 &  1.630 &      1.953 &      2.912 &      0.732 &      0.518 &      0.234 &      0.419 &      2.003 \\
1999 & 4.548 &  2.252 &      2.543 &      2.947 &      1.064 &      0.600 &      0.275 &      0.868 &      2.876 \\
2000 & 4.685 &  2.790 &      2.822 &      2.241 &      1.051 &      0.410 &      1.189 &      1.028 &      3.604 \\
2001 & 4.681 &  2.281 &      5.921 &      2.348 &      1.242 &      0.553 &      0.274 &      0.951 &      3.018 \\
2002 & 4.671 &  3.781 &      6.187 &      3.108 &      1.507 &      0.786 &      0.383 &      0.898 &      5.119 \\
2003 & 4.993 &  3.843 &      6.604 &      3.866 &      1.618 &      1.199 &      0.522 &      1.057 &      5.647 \\
2004 & 5.238 &  3.694 &      6.237 &      2.171 &      1.500 &      0.906 &      0.597 &      1.158 &      5.841 \\
2005 & 5.352 &  4.223 &      6.308 &      1.921 &      1.225 &      0.871 &      0.495 &      1.735 &      5.377 \\
2006 & 5.057 &  3.971 &      6.119 &      2.220 &      1.651 &      1.399 &      0.771 &      1.588 &      4.854 \\
2007 & 5.249 &  4.259 &      6.405 &      1.714 &      1.870 &      1.461 &      0.834 &      1.390 &      5.019 \\
2008 & 5.185 &  4.153 &      6.331 &      1.784 &      1.741 &      1.261 &      1.283 &      2.564 &      4.769 \\
2009 & 5.103 &  4.179 &      7.364 &      1.675 &      1.046 &      1.186 &      1.404 &      3.786 &      4.481 \\
2010 & 4.888 &  4.425 &      6.063 &      1.632 &      1.109 &      0.842 &      1.437 &      1.590 &      4.555 \\
2011 & 4.900 &  4.587 &      6.024 &      1.411 &      1.183 &      1.012 &      1.686 &      2.259 &      6.024 \\
2012 & 5.521 &  5.084 &      6.733 &      1.850 &      1.030 &      1.399 &      2.064 &      3.120 &      4.965 \\
\hline
Average&4.935&  3.677 &      5.574 &      2.253 &      1.305 &      0.960 &      0.897 &      1.627 &      4.543 \\
\hline
\end{tabular}
\\
\\
{\bf Note:} {\em (MNRAS) Montly Notices of the Royal Astronomical Society; (A\&A) Astronomy and Astrophysics, 
(ApJ) Astrophysical Journal, (NewA) New Astronomy, (IJMPD) International Journal of Modern Physics D,
(AN)  Astronomische Nachrichten, (Ap\&SS) Astrophysics \& Space Science, (PASA) Publications of the 
Astronomical Society of Australia, (AJ) Astronomical Journal.} 
\end{table}

\subsection{Application of the Institutional Performance Score to Turkish Institutions}

We present in Table 2, the resulting institutional performance scores ({\em IPS}) of nine 
leading Turkish institutions, along with their number of publications, each component 
of the {\em IPS}, namely $IF \times AC$ (represented with \textcircled{1} in Table 2) and 
$(n_{citations}/n_{years}) \times AC$ (\textcircled{2} in Table 2), as well as their 
institutional {\em h}- and the other indices. The table is formed in such a way that 
the upper part is for all 749 publications, and the lower part is formed by considering 
564 publications whose leading author reside in Turkey. The institutes in both portions 
of Table 2 are ranked according to their {\em IPS} values. 

\begin{table}
\setlength{\tabcolsep}{4pt}
 \renewcommand{\arraystretch}{.7}
\caption{The list of nine leading institutions with their number of papers 
($N$), total author contribution corrected journal impact ($\textcircled{1}$), 
author contribution corrected individual impact (\textcircled{2}), the ratio 
of $\textcircled{2}$ to $\textcircled{1}$, the {\em IPS} for all publications 
(upper portion) and for the publications with leading authors based in Turkey 
(lower portion). We also list few other performance indicators, namely 
{\em h}-index, {\em g}-index, {\em AR}-index, and $IF^2$.}
\begin{tabular}{clccccccccc}
\hline
Rank & Institution & $N$ & $\textcircled{1}$ & \textcircled{2} & $\textcircled{2}/\textcircled{1}$ & {\em IPS} & {\em h}-index & {\em g}-index &  {\em AR}-index & $IF^2$ \\
\hline
         1 &   Sabanc\i &        105 &        216 &        115 &       0.53 & 3.15 & 24 &         35 &      12.45 &         40 \\
         2 &        Ege &        135 &        305 &        104 &       0.34 & 3.03 & 15 &         23 &       7.63 &         27 \\
         3 &       BOUN &         62 &        106 &         32 &       0.30 & 2.23 & 12 &         18 &       5.18 &         17 \\
         4 &       COMU &        115 &        180 &         72 &       0.40 & 2.19 & 18 &         25 &       8.74 &         26 \\
         5 & \.Istanbul &        122 &        170 &         88 &       0.52 & 2.11 & 20 &         28 &       8.86 &         27 \\
         6 &       METU &        120 &        191 &         48 &       0.25 & 1.99 & 20 &         33 &       11.6 &         28 \\
     7 &T{\"U}B{\.I}TAK &         62 &         80 &         15 &       0.19 & 1.53 & 12 &         18 &       7.03 &         21 \\
         8 &     Ankara &         71 &         75 &         31 &       0.41 & 1.49 & 11 &         23 &       8.64 &         15 \\
         9 &    Akdeniz &         45 &         45 &         16 &       0.36 & 1.36 & 11 &         15 &       5.15 &         17 \\
\\
\hline
Rank & Institution & $N$ & $\textcircled{1}$ & \textcircled{2} & $\textcircled{2}/\textcircled{1}$ & {\em IPS} & {\em h}-index & {\em g}-index &  {\em AR}-index & $IF^2$ \\
\hline
         1 &   Sabanc\i &         43 &        140 &         76 &       0.54 & 5.02 & 14 &         23 &       8.21 &         31 \\
         2 &        Ege &         93 &        255 &         85 &       0.33 & 3.66 & 11 &         18 &       6.07 &         25 \\
         3 & \.Istanbul &         62 &        115 &         68 &       0.59 & 2.95 & 14 &         19 &       6.54 &         20 \\
         4 &       METU &         66 &        151 &         32 &       0.21 & 2.77 & 13 &         21 &       5.98 &         24 \\
         5 &       COMU &         82 &        159 &         59 &       0.37 & 2.66 & 11 &         16 &       5.49 &         19 \\
         6 &       BOUN &         41 &         83 &         25 &       0.30 & 2.63 &  9 &         15 &       4.62 &         11 \\
    7 & T{\"U}B{\.I}TAK &         31 &         63 &         11 &       0.17 & 2.39 &  7 &         11 &       3.10 &         15 \\
         8 &    Akdeniz &         24 &         33 &         13 &       0.39 & 1.92 &  6 &          8 &       3.77 &          9 \\
         9 &     Ankara &         40 &         52 &         15 &       0.29 & 1.68 &  6 &         10 &       2.80 &          8 \\
\hline
\end{tabular}  
\end{table}

We find that Sabanc\i~University appears on top of the list in both publication 
categories, followed by Ege, which produced the largest number of institutional 
publications in our sample. It is noteworthy that Ege University was founded in 
1962 while Sabanc\i~University in 1999 and the average number of researchers in 
Ege have been much larger than that in Sabanc\i~University. Our proposed 
performance indicator is not biased by such contrasts since we normalize the 
total quantities by the number of papers published. With respect to the 
individual impact (\textcircled{2} in Table 2), Sabanc\i~again earns the 
first rank, followed by Ege and \.Istanbul. Another important ranking tool here 
is the ratio of \textcircled{2} to \textcircled{1}, that is the fraction of the 
collective impact of scientific papers within the collective impact they gained 
by their respective journals. In this scheme, Sabanc\i~leads, and closely 
followed by \.Istanbul University.

Sabanc\i~University ranks on top also in other performance indicators. It is striking to 
note in Table 2 that METU, \.Istanbul and COMU the second, third and fourth 
places, respectively, in their respective {\em h}-index \citep{Hirsch05}, 
{\em g}-index \citep{Egghe06}, {\em AR}-index \citep{Jin07}, and $IF^2$ 
\citep{Boell10} rankings. In the {\em IPS} ranking, METU, \.Istanbul and COMU rank 
sixth, fifth and fourth, respectively.

\begin{figure}
\centering
   \includegraphics[scale=0.60]{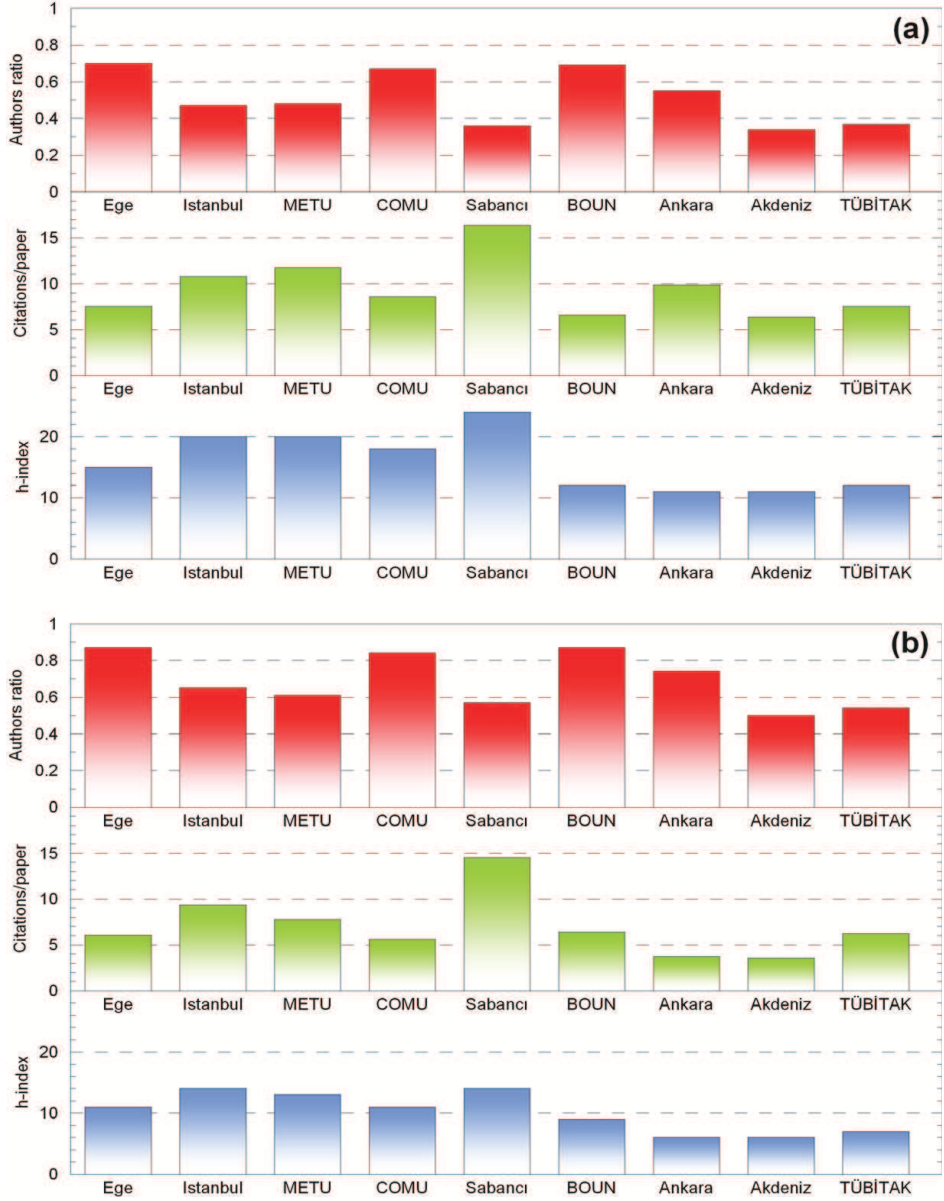}
   \caption{Average author contributions, citations per papers and {\em h}-index of 
institutions for all papers (a) and for papers with domestic leading authors (b).}
\end{figure}

We present in Fig. 2, the average author contribution ($AC$) for each 
institution in both publication groups. We find that the average author 
contribution ratios from Turkish institutions to all publications vary between 
0.34 and 0.70. For all 749 publications, Turkey resident author contribution to 
research papers mostly comes from BOUN, Ege and COMU (Fig. 2a). There are four 
institutions that were found to pass the author contribution ratios of 0.50, while 
Sabanc\i~University remains below this proportion, even though with the highest 
citation value received for research papers. The author contribution ratios for 
publications with domestic leading authors varied between 0.50 and 0.87. In this 
category, BOUN and Ege earn the first place, followed by COMU with 0.84 and Ankara  
University with 0.74. Also in Fig. 2, we present the number of citations per papers 
and institutional {\em h}-indices of all nine Turkish institutions. When citations 
per papers were considered, Sabanc\i~University leads in both publication 
categories with 16.36 and 14.53 citations per paper, respectively. It is followed 
by METU (11.74\%) and \.Istanbul (10.78\%) for all publications and \.Istanbul 
(9.37\%) and METU (7.76\%) for publications with domestic leading authors. 

Finally, we construct time evolution of {\em IPS} values of Turkish astronomy and 
astrophysics related research institutions with exceeding 100 publications. We also
calculate {\em h}-index, {\em g}-index, {\em AR}-index, and $IF^2$ for these five
institutions to compare with our proposed performance indicator. As seen in Fig. 3, 
the annual {\em IPS} of Sabanc\i~University is mostly in the 4-8 band over the course of our 
study from 1998 to 2012. Note the fact that Sabanc\i~is a newly established 
institution and astrophysical research started in 1999. Ege, METU and \.Istanbul 
Universities lie around {\em IPS}s of 4. It is noteworthy that Ege exhibits a 
gradual increase trend until 2004. The {\em IPS} trends of other institutions appear 
between 2 and 4. It is important to note that {\em h}-index, {\em g}-index, 
{\em AR}-index, and $IF^2$ exhibit cumulative evolution in time, while the 
{\em IPS} can evolve positive or negatively, depending on the scientific impact of 
research units.

\begin{figure}
\centering
   \includegraphics[scale=0.63]{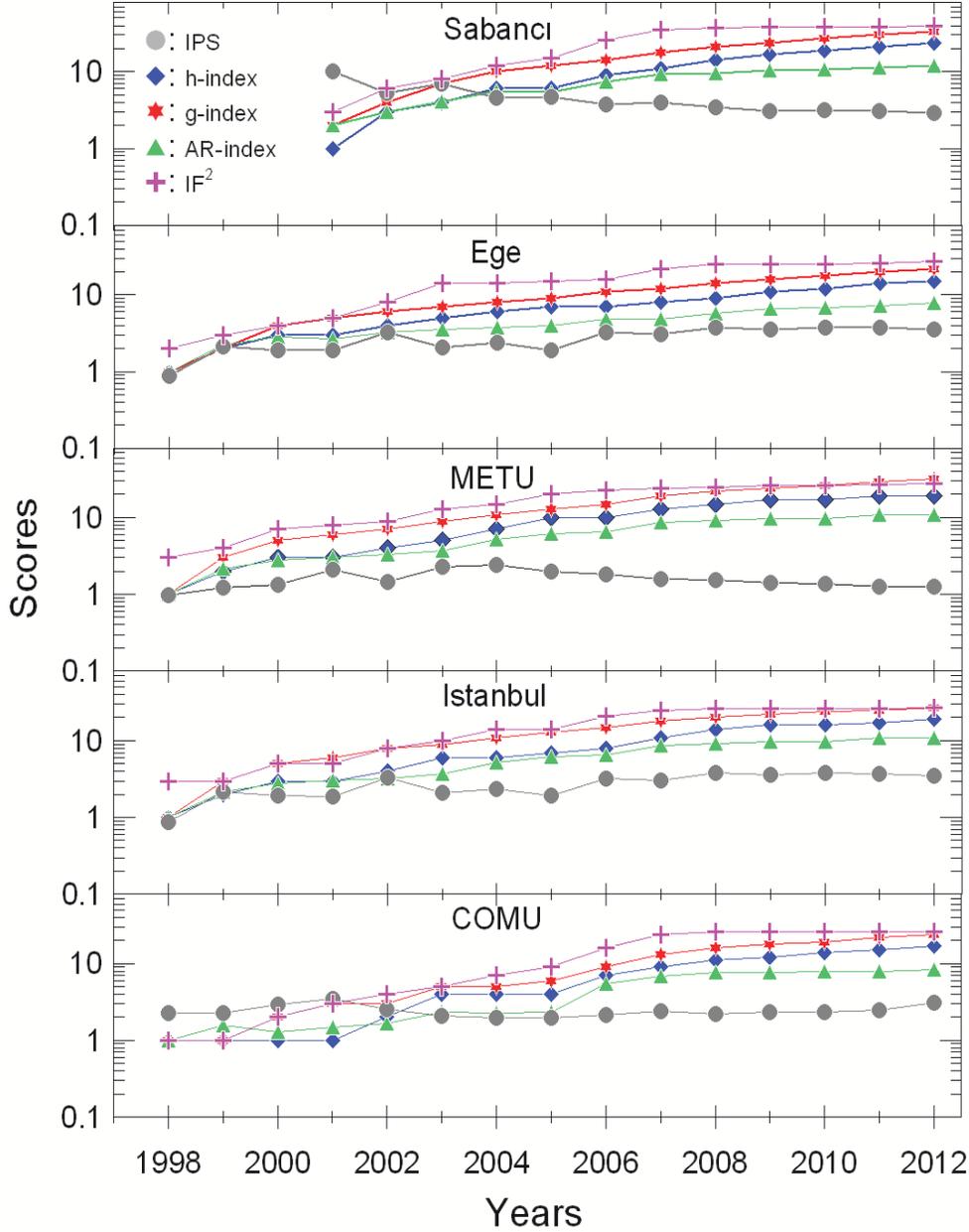}
   \caption{Evolution of {\em IPS}, {\em h}-index, {\em g}-index, {\em AR}-index, 
and $IF^2$ for five institutions with more than 100 publications.}
\end{figure}

\section{Discussion}
We introduced a new quantitative indicator to evaluate scientific performance of 
research institutions. Our proposed indicator consists of two crucial components: 
one is the author contribution weighted impact factor of the journal in which a 
paper has been published, and the other one is again author contribution corrected 
the number of citations received by the paper per each year since its appearance in 
the journal. In other words, the {\em IPS} value can be regarded as institutional scientific
impact of a research unit.

In the era of very high-speed communications and rather easy access to 
high-performance computation, scientists of today are greatly benefiting from the 
fact that geopolitical borders are no longer boundaries for scientific 
collaborations. As reaching out for international collaborations gets easier, 
the sizes of international research teams become eventually larger. When it comes 
to extensively large experimental efforts, such as, the Large Hadron Collider 
project at CERN, the size of collaborations can be as large as thousands of 
researchers from hundreds of different institutions. Therefore, it would not be 
trivial to assess the outcome of their collaborative effort (peer-reviewed papers) 
to a particular institution only. For this reason, we include the ratio of the number 
of co-authors from a particular institution to the total number of co-authors as a 
multiplicative weight for the impact factor of the journal in which a particular 
paper has been published.

The journal, and its associated impact factor cannot provide a direct measure for 
the quality of a research topic. Some articles might end up in a journal with no 
page charge but has a low impact factor due to the lack of funding for publication 
costs. Nevertheless, there are, fortunately, numerous journals which require no 
publication charges but have high impact factors, such as MNRAS in the field of 
astronomy and astrophysics. When folded with the ratio of contributing authors, 
the impact factor becomes a more sensitive quality indicator of a research paper.

Another important achievement indicator of a scientific paper is the number of 
citations received. It is unavoidable that a paper takes some time for its 
visibility before it is being referred by peer-researchers. As years pass by, it 
will be eligible for further referral. In our parameterization, we consider the 
citation based impact of a paper per the number of years passed so that the outcome 
is balanced for newly published papers, as well as those published a while ago 
and had already ample periods of time for their visibility.

We apply our new performance indicator scheme to the outputs of Turkish institution 
specialized in astronomy and astrophysics. We clearly find that commonly used {\em h}-index 
or its variants suggest slightly different rankings for the same sample since they 
involve primarily citations received by papers. This approach underestimates the 
performance of an institution which produced modest number of highly cited papers. 
As we showed in Table 2, {\em h}-index, {\em g}-index, {\em AR}-index, and $IF^2$
based ranking closely resemble each other. On the other hand, the {\em IPS} ranking
is significantly different. Another important property of the use of {\em IPS} is that
it can grow or decay, depending on the scientific performance of research
institutes. Whereas, the other four indicators compared here evolve in time
cumulatively.

We also investigated other proposed performance indicators \citep[such as][]
{Batista06, Vieira10, Abramo13, Franceschini13}. The indicators proposed by \citet{Abramo13} 
requires the number of full time equivalent staff of research institutions, which 
is, in most cases, not easy to obtain for the  institutions other than the home 
institution of a researcher. The methods proposed by \citet{Batista06} and 
\citet{Vieira10} differ from pure {\em h}-index analysis, but still heavily based 
upon {\em h}-index parameters. Our proposed scheme, on the other hand, makes use of 
easily available input parameters, which can be extracted from various commonly 
used channels, such as, Web of Knowledge.  

Finally, it is important to note that our proposed performance indicator can also 
be adopted to evaluate scientific output of an individual researcher. For this 
purpose, the weighting parameter of the journal impact factor (i.e. the author 
contribution ratio) is simply replaced with the reciprocal of the number of 
co-authors. When summed over all publication of researchers, this would provide 
a more sensitive comparison tool for personal evaluations.

\end{document}